\documentclass{ifacconf}

\usepackage{graphicx}      
\usepackage{natbib}        
\usepackage{amsmath,amssymb,amsfonts}

\usepackage{algorithm}
\usepackage{algpseudocode}
\usepackage{graphicx}
\usepackage{textcomp}
\usepackage{multirow}
\usepackage{subfigure}
\usepackage{gensymb}
\usepackage{url}

\usepackage[dvipsnames]{xcolor}

\usepackage{array}
\begin{document}
\begin{frontmatter}

\title{Nonlinear Moving Horizon Estimation and Model Predictive Control for Buildings with Unknown HVAC Dynamics\thanksref{footnoteinfo}} 

\thanks[footnoteinfo]{Saman Mostafavi and David Schwartz are with the Palo Alto Research Center, Inc. (PARC). Harish Doddi is currently with NVIDIA Corp. This work was performed while all the authors were affiliated with PARC.\\
\\
Accepted to 4th IFAC Workshop on Cyber-Physical and Human Systems (CPHS 2022)}%

\author[First]{Saman Mostafavi} 
\author[First]{Harish Doddi} 
\author[Second]{Krishna Kalyanam} 
\author[First]{David Schwartz} 

\address[First]{Palo Alto Research Center, Inc. (PARC), 
   Palo Alto, CA 94304 USA (e-mail: \{smostafa, dschwart\}@parc.com).}
\address[Second]{NASA Ames Research Center, 
   Moffett Field, CA 94035 USA (e-mail: krishnak@ucla.edu)}

\begin{abstract}                
We present a solution for modeling and online identification for heating, ventilation, and air conditioning (HVAC) control in buildings. Our approach comprises: (a) a resistance-capacitance (RC) model based on first order energy balance for deriving the zone temperature dynamics, and (b) a neural network for modeling HVAC dynamics. State estimation and model identification are simultaneously performed using nonlinear moving horizon estimation (MHE) with physical constraints for system states. We leverage the identified model in model predictive control (MPC) for occupant comfort satisfaction and HVAC energy savings and verify the approach using simulations. Our system relies only on building management system data, does not require extensive data storage, and does not require a detailed building model. This can significantly aid the large scale adoption of MPC for future occupant-centric control of grid-interactive buildings.
\end{abstract}


\end{frontmatter}

\section{Introduction}
According to recent estimates by \cite{doe/eia}, residential and commercial buildings account for nearly 40\% of energy usage in the United States. Most of this energy provides services to occupants, the greatest chunk of which can be associated with cooling and heating of spaces for occupant comfort and satisfaction. Better management of HVAC control systems can improve indoor thermal conditions and lead to more pleasant living spaces, better working conditions and safer environments. Specific examples include better productivity in offices \citep{seppanen2004control}, better learning performance and pupil attendance in classrooms \citep{wargocki2020relationships}, and more recently, mitigation of transmission of COVID-19 in office buildings and hospitals. These can be achieved through a combination of heating and cooling for the indoor environment, more outdoor air circulation, and stricter standards for space pressure control, but may come at the cost of increasing the energy consumption for buildings and contributing to more carbon emission and less energy efficiency.

The problem quickly scales up for large number of buildings where HVAC control system can play the part of a variable load in a coordinated and integrated grid. To mitigate and manage, advanced controls, implemented as a supervisory control system, can be used to serve multiple end-use cases such as load shifting, demand response, and islanding \citep{satchwell2021national}. Model predictive control (MPC) is a particularly powerful approach for handling hard constraints for state and control inputs in nonlinear multivariable control systems. In simulation studies as well as field studies without time or budget constraints on the modeling process, MPC is the standard for comfort-oriented and energy-efficient building control, as long as a model of the plant exists or can be formulated and identified \citep{sturzenegger2013model,kelman2011bilinear,zeng2019identification}. In reality, developing a control model for the uncertain built environment is the main obstacle for deployment of MPC, accounting for as much as 70\% of the total effort of setting up MPC controllers \citep{atam2016control}. Challenges including heterogeneity of components, uncertainty in data and model, and the size and complexity of buildings significantly limit the potential of a scalable formalism for detailed high-fidelity physics-based model generation. First order models are common alternative and are much easier to formulate, however, they need constant calibration to accommodate inaccuracies associated with significant lumping of the true nonlinear dynamics. Recent studies approach the modeling challenges in various ways such as modular model development\citep{sturzenegger2014brcm}, differentiable controls \citep{drgonadifferentiable} and using Gaussian Process for chance constrained MPC \citep{jain2018learning}. 

The focus of this paper is the modeling and identification of a control-oriented building model comprising a gray-box lumped-parameter zone model and augmented with a neural network for unknown HVAC dynamics. For each time period, we perform parameter and state estimation and update the model to be used in the MPC controller. The first computed optimal input from MPC is then applied to the building HVAC system and the measurements are collected for the next iteration of estimation and control. Although we are aware of other linear time-varying (LTV) MPC approaches \citep{rastegarpour2019experimental}, to the best of our knowledge, this is the first study to consider a sequential MHE/MPC algorithm for identification of nonlinear time-varying models and control in building energy optimization. 
We present the details and demonstration of the approach in the following order. In Section II, we provide a detailed summary of the modeling methodology, Section III outlines the formulation of the online identification and MPC algorithms. Section IV demonstrates the effectiveness of the proposed approach in simulation and Section V concludes the paper with remarks regarding future research directions. 

\section{System Model}
\begin{figure}[b!]
  \centering
  \includegraphics[width=0.9\columnwidth]{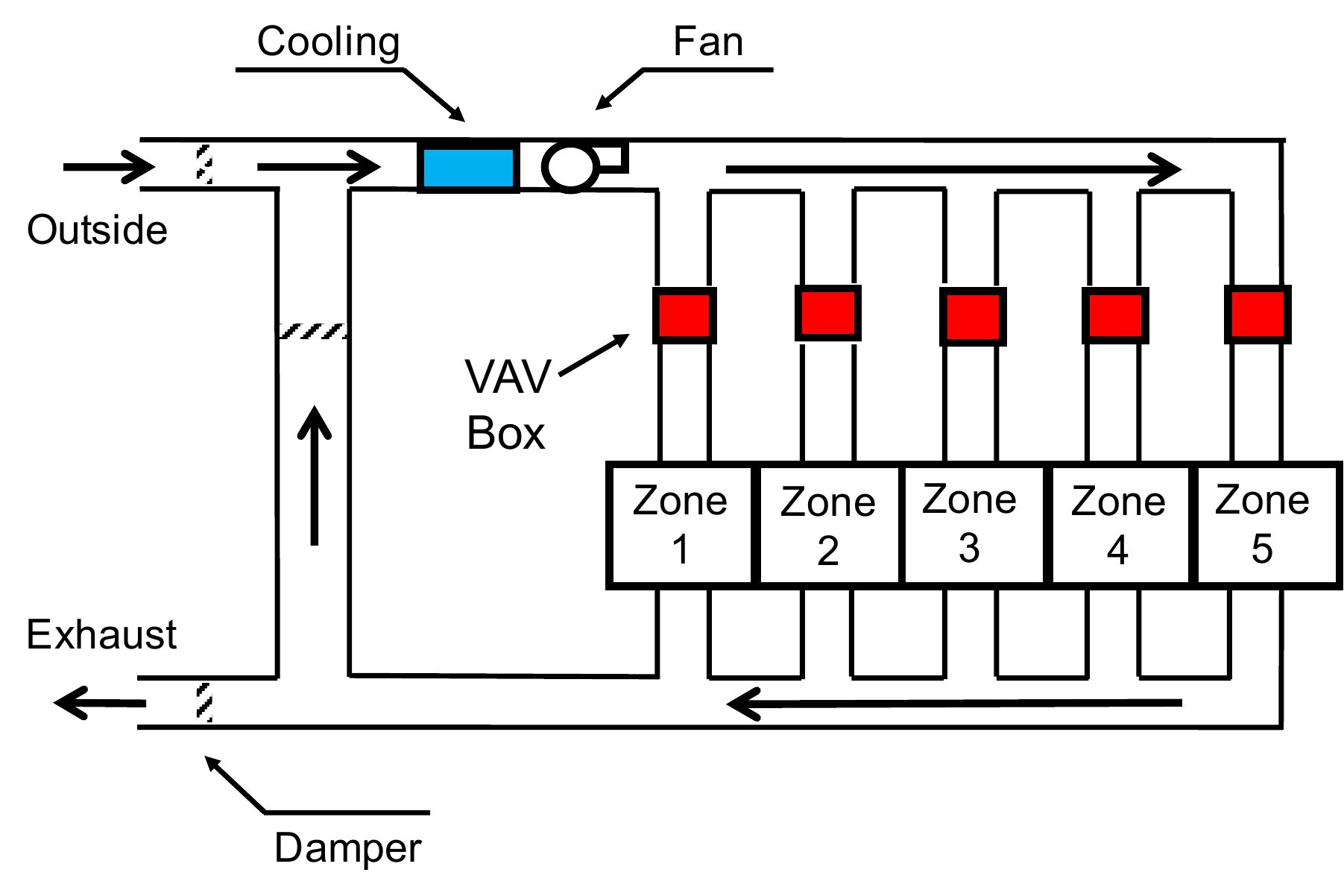}
  \caption{Overall schematic of studied HVAC system.}
  \label{fig:overal_HVAC}
\end{figure}

Figure \ref{fig:overal_HVAC} depicts the schematic of the model building and HVAC system. Highlighted in blue is the central cooling system that regulates the  supply air temperature. A supply fan controls the total airflow rate. The central chilled supply air is divided equally between five separate zones and is fed to each individual via a variable air volume (VAV) reheat box, consisting of a heating unit and dampers. The reheat box controls the supply temperature and flow rate to comply with the thermostat set points of each individual zone. The return air from each individual zone is mixed with outside air to achieve ventilation requirements. Similar to \cite{kelman2011bilinear} we neglect humidity and use a first order RC model to track average zone-level temperatures. Due to heterogeneity of HVAC equipment, one cannot effectively generalize a physics-based lower-order structure for modeling supply air temperature dynamics. Therefore, we treat the dynamics of HVAC as unknown and model them with a neural network. The details of the modeling approach are provided below.  

\subsection{Modeling the dynamics for zone temperatures} \label{sec:Thermal_zone}
We consider a thermal zone $i$ with temperature $T_{z,i}$ and thermal capacitance $C_i.$ $T_{z,j}$ represents the temperature of a neighboring thermal zone $j$ that is separated by a wall of thermal resistance $R_{ij}$. Applying Kirchhoff's law, in continuous time the total heat load on a thermal node $i$ for $n$ zone building can be expressed via the following differential equations:
\small
\begin{subequations}
\label{eqn:gov_ZoneTemp}
\begin{align}
&C_{i}\frac{dT_{z,i}}{dt} = \sum\limits_{j=1,j\neq{i}}^{n+1} a_{ij}\frac{T_{z,j}-T_{z,i}}{R_{ij}}+ \dot{Q}_{tot,i} \\
&\dot{Q}_{tot,i} = \dot{Q}_{occ,i}+ \dot{Q}_{sol,i}+\dot{Q}_{hvac,i}+\dot{Q}_{other,i} \\
&\dot{Q}_{hvac,i} = m_{z,i}c_{pa}(T_{supp,h,i}-T_{z,i}), \\
&\dot{Q}_{sol,i} = A^i_w\eta_{sol}
\end{align}
\end{subequations}
\normalsize
where $a_{ij}=1$ if there exist a wall between zone $i$ and zone $j,$ else $a_{ij}=0$. Note that for $n$ zones, we have $n+1$ thermal nodes where the last one is the measured ambient temperature. Total thermal current comprises of $\dot{Q}_{occ,i}$ representing occupant thermal load, $\dot{Q}_{sol,i}$ solar irradiation,  $\dot{Q}_{hvac,i}$ supply air entering the room and $\dot{Q}_{other,i}$ other stochastic disturbances (equivalent to a process noise in state space form). $T_{supp,h,i}$ represents the supply air temperature entering the zone $i$ with specific heat capacity $c_{pa},$ and $A^i_{w}$, $\eta_{sol}$ are  window area for zone $i$ and measured solar irradiance respectively. It is important to note that we assume occupancy schedules and weather forecast for the control horizon are available on demand through occupancy sensor data, building schedule and online weather services. The zone model has the parameters $C_i,R_{i,j},A^i_w~\text{for}~i \in \{1,2,...,n\}$  to be learned from data.




\subsection{Modeling the dynamics for supply air temperature}\label{sec:Heat_Exchanger}

It is typical for commercial HVAC systems to have a central unit supplying air to several distributed components. Without loss of generality, we can also assume that each of these components are controlled by fine tuned PI controllers as is common practice in commercial buildings. Given the heterogeneity of HVAC systems, we consider a nonlinear data driven model, namely a neural network model that can be scaled to any type of HVAC system. Looking at Figure~\ref{fig:overal_HVAC}, the supply air with temperature $T_{supp,c}$ is distributed to all the zones, where it is reheated to $T_{supp,h,i}$ by a VAV reheat box associated with zone $i$. The aim of the data driven model is to have good predictive accuracy for  $T_{supp,c}, T_{supp,h,i}$ for a control horizon of interest. 
The dynamics of cooling supply is modeled by the following feed forward Neural Network $f_{N,c}$ with parameters $\theta_{N,c}$:
\small
\begin{equation}
\label{eqn:fnn}
\begin{aligned} 
    &T_{supp,c}(t+\Delta t) = f_{N,c}(T_{mix}^{(t-M:t)},T_{supp,c}^{(t-M:t)},u^{(t-M:t)},\theta^{(t)}_{N,c}) \\ 
\end{aligned}
\end{equation}
\normalsize

where $T_{mix}$ is the mixed air temperature at the inlet of cooling system and manipulated variable $u$ is the cooling setpoint for supply air $T_{cs}$. Assuming no heat loss or gain in the ducts, the mixed air temperature is calculated as:
\small
\begin{subequations}
\label{eqn:mix_air}
\begin{align} 
    &T_{mix}=r_{mix}T_{amb}+(1-r_{mix})T_{z,avg} \\ 
    &T_{z,avg}= \frac{1}{n}\sum\limits_{i=1}^n T_{z,i}
\end{align}
\end{subequations} 
\normalsize
from all the room exhausts $T_{z,avg}$ plus outside air given some mixing ratio $r_{mix}$. 

Notice that in (\ref{eqn:fnn}), $(T_{mix}^{(t-M:t)},T_{supp,c}^{(t-M:t)},u^{(t-M:t)})$ represents the $M+1$ lags of states and manipulated variables that are used for the prediction of the next state along with $\theta^{(t)}_{N,c}$ , the NN parameter estimates at time $t$. This formulation is related to our choice of estimator, i.e., moving horizon estimator, for model identification. Similarly, for the VAV reheat box, we consider the following NN for $T_{supp,h,i}$ prediction:
\small
\begin{equation}
\label{eqn:fnn2}
\begin{aligned} 
    &T_{supp,h,i}(t+\Delta t) = f_{N,i}(T_{supp,h,i}^{(t-M:t)},T_{supp,c}^{(t-M:t)}, u^{(t-M:t)}_{i},\theta^{(t)}_{N,i}) \\ 
\end{aligned}
\end{equation}
\normalsize
$\forall i \in \{1,2,3,...,n\}$ where manipulated variable $u$ represents individual zone set points $T_{zs,i}$ and $f_{N,i}$ represents neural networks that learn the dynamics for VAV boxes. Similar to the cooling system, we use the $M+1$ lags of control inputs and state in our model. For generalizability, it should be noted that it is a reasonable assumption to expect the availability of all these data points with a frequency of once per five minutes or better through the building management system. The final HVAC model has parameters $\{\theta_{N,c},\theta_{N,i}\}$, $\forall i \in \{1,2,3,...,n\}$ to be learned from data.

\section{ONLINE IDENTIFICATION AND MODEL PREDICTIVE CONTROL}
We consider the following continuous time nonlinear system model that abstracts equation (1-4): 
\begin{subequations}
\label{eqn:cont_time}
\begin{align}
   &\dot{x}(t)=f(x(t),u(t),p(t),p_{\text{tv}}(t))+w(t) \\ \nonumber
   &x(t_0) = x_0 \\ \nonumber
   &y(t)=h(x(t),u(t),p(t),p_{\text{tv}}(t))+v(t)\\ \nonumber
    \text{s.t.}\\
    & T^{l}_{supp,c} \le T_{supp,c}\le T^{u}_{supp,c} \\
    & T^{l}_{supp,h,i} \le T_{supp,h,i}\le T^{u}_{supp,h,i} \\
    & T_{heat,i}(t) \le T_{z,i}(t) \le T_{cool,i}(t)\\
    & T_{supp,c}\le T_{mix}\\
    & T_{supp,c}\le T_{supp,h,i}
\end{align}
\end{subequations}
Here, $p_{\text{tv}}$ are exogenous disturbances $\{\eta_{sol},T_{amb},\dot{Q}_{occ}\}$, $p_t$ are model parameters $\{\theta_{N,i},\theta_{N},\theta_{RC}\}$ to be learned from data, and system states $x$ and manipulated variables $u$ are concatenation of (1-4) as described in the previous section. $w(t)$, $v(t)$ represent unmodelled system disturbances and measurement noise assumed to be zero mean, Gaussian, white and stationary. Constraints are interpreted as follows: a desired cooling supply air and desired VAV reheat supply air temperature range (5b-5c), desired comfort range for individual zones (5d) and strict cooling and heating for the cooling and heating units respectively (5e-5f).

\subsection{Online model identification}
We adopt a moving horizon strategy similar to one described in \cite{kuhl2011real} since MHE allows for nonlinear modeling of the system dynamics, allows imposing physical constraints on states, and also allows an integration of uncertainties. The algorithm is a recursive estimation technique where for a finite length sliding window of size $M+1$ the estimates $\hat{x}_{(t-M)},...,\hat{x}_{(t)} \text{,} \hat{p}_t$ are obtained from previous measurements and control inputs ${y}_{(t-M)},...,y_{(t)} \text{,} {u}_{(t-M)},...,u_{(t)}$. We use direct collocation \citep{andersson2019casadi} which parameterizes the entire state trajectory as piecewise low-order polynomials and include these as decision variables in the optimization. 

Considering that, we now discuss the joint state and parameter estimation formulation. At time $k,$ given the past state estimates $\boldsymbol{x}_o^k=[\hat{x}^{(k-M-1)^T}, \cdots, {\hat{x}^{(k-1)^T}}]^T,$ the parameter estimate $\hat{p}^{k-1},$ measurements $\{ y(k+l) \}_{l=-M}^{0}$ and control inputs $\{u(k+l) \}_{l=-M}^{0},$ the MHE solves for $\{x_{o,*}^{k-M} \boldsymbol{x}_*^{k}, p_*^k \}:=$
\small
\begin{equation}
\label{eqn:MHE}
\begin{aligned}
    &arg \min_{ \{ \Tilde{x}_o, \boldsymbol{\Tilde{x}}^{k}, \Tilde{p}^k \} }~~~ \left\|\Tilde{x}_o - \hat{x}^{(k-M-1)} \right\|^2_{P_x} + \left\|\Tilde{p}^k - \hat{p}^{k-1} \right\|_{P_p}+\\
    &  \sum\limits_{t=k-M-1}^{k} \Big[ \left\|y(t)-h( \boldsymbol{\Tilde{x}}^k,u(t),\Tilde{p}^k,p_{\text{tv}}(t))\right\|^2_{P_v}+\left\|w(t)\right\|^2_{P_w} \Big]\\
    &s.t.~~ \hat{\boldsymbol{X} } = \{ \Tilde{x}_o, \boldsymbol{\Tilde{x}}^{k} \in \mathbb{R}^{M}, \Tilde{p}^k \} \in \Omega,
\end{aligned}
\end{equation}
\normalsize



\noindent where, $\|l\|_M^2:= l^TMl$ is the weighted vector norm, $P_x,P_p.P_v$ and $P_w$ are symmetric, positive semi-definite matrices with appropriate dimensions. In particular, $P_v$ and $P_w$ are inverse of covariance matrices for measurement and process noise and penalize the state and measurement discrepancies. Optimal solution (\ref{eqn:MHE}) provides the state and parameter estimate for time $k$ as follows: $[\hat{x}^{(k-M)^T}, \cdots, {\hat{x}^{(k)^T}}]^T= \boldsymbol{x}_o^{k+1} = [x_{o,*}^{(k-M)^T}, \boldsymbol{x}_*^k]$ and $\hat{p}^k = p_*^k.$ The feasible set $\Omega$ imposes the system dynamics and constraints in (\ref{eqn:cont_time}). The parameter set $p$ consist of all model parameters for neural networks and RC models of equations (\ref{eqn:gov_ZoneTemp}), (\ref{eqn:fnn}) and  (\ref{eqn:fnn2}).
The moving horizon approach curtails the curse of dimensionality by reducing the degrees of freedom in the constrained least-square optimization in (\ref{eqn:MHE}) (as opposed to full-information estimator). In (\ref{eqn:MHE}), we consider states and model parameters as free parameters. This estimation results in time-varying model parameters for which, we regularize the parameter noise with $P_p$. For more details about MHE the readers are referred to \cite{kuhl2011real}. 

The proposed MHE algorithm is computed as follows:
\begin{enumerate}
    \item Tune weight matrices $P_x,P_p.P_v$ and $P_w$ on historical data. 
    \item Given the $\boldsymbol{x}_o^k=[\hat{x}^{(k-M-1)^T}, \cdots, {\hat{x}^{(k-1)^T}}]^T,$ the parameter estimate $\hat{p}^{k-1},$ measurements $\{ y(k+l) \}_{l=0}^{-M}$ and $\{u(k+l) \}_{l=0}^{-M},$ MHE solves for $\{x_{o,*}^{k-M} \boldsymbol{x}_*^{k}, p_*^k \}$ using (\ref{eqn:MHE}) 
    \item $[\hat{x}^{(k-M)^T}, \cdots, {\hat{x}^{(k)^T}}]^T= \boldsymbol{x}_o^{k+1} = [x_{o,*}^{(k-M)^T}, \boldsymbol{x}_*^k]$ and $\hat{p}^k = p_*^k.$
    \item $k=k+1$ and repeat steps $2$ and $3$.
\end{enumerate}

\subsection{MPC Formulation}
MPC is commonly used for setpoint tracking by minimizing deviations. In this work, we focus on a comfort oriented energy minimization that also adheres to operational bounds. In addition, MPC is assumed to take a supervisory controller role for lower level control systems in the central cooling and heating units where at each time step, MPC generates tracking setpoints for lower level controllers. This formulation is particularly appealing for retrofitting legacy fine tuned control systems that are already in operation in commercial buildings. More details are provided in the result section. Next, we define the cost, i.e., HVAC power consumption, and constraints associated with the formulation of energy optimization problem.

\textit{Power consumption} 
For the purpose of this study, the energy consumers in HVAC thermal cooling and heating costs in the building are: (i) central cooling unit of HVAC system, (ii) individual VAV reheat boxes of the zones\footnote{The mass flow rates of the air ($m_i, m_{z,i}$) in the air loop is kept constant and hence the supply fan power is not considered in the MPC objective.}. We consider the energy consumption of cooling (heating) units to be a function of of cooling (heating) load of the air. Suppose $P_c,\ P_h$ are the cooling coil power and the heating coil power, respectively. Then, at time $k$:
\small
\begin{subequations}
\label{eqn:power}
\begin{align}
&P_c^k = \frac{1}{\eta_c}(\sum_i m_i)c_{pw}[T_{mix}^k-T_{supp,c}^k],\\
&P_h^k = \frac{1}{\eta_h}\sum_i m_{z,i}c_{pa}[T_{supp,h,i}^k- T_{supp,c}^k].
\end{align}
\end{subequations}
\normalsize
From (\ref{eqn:cont_time}f) and (\ref{eqn:cont_time}g), it follows that $P_c^k$ and $P_h^k$ are non-negative.

\textit{Measurements ($y$):} We assume the following states as measured: $\{T_{supp,c}, T_{supp,h,i}, T_{z,i} \  \forall i \}.$ In addition, we assume the following forecast for disturbances for the duration of the prediction horizon: $\{\eta_{sol}, T_{amb}, \dot{Q}_{occ}\}$ via online weather services and pre-determined building operation schedules.  

\textit{Time varying constraints:} 
The constraints are provided by the system dynamics as shown in (\ref{eqn:cont_time}a-\ref{eqn:cont_time}f). Specifically, constraint (\ref{eqn:cont_time}d) satisfies occupant comfort in zone $i$, i.e., occupant is considerd to be comfortable when $T_{z,i}(t) \in [T_{heat,i}(t),T_{cool,i}(t)]$ at any time $t.$

\textit{MPC Objective:}
We implement the control algorithm in a receding horizon manner where an optimization over $N$ time steps is solved at each step $k$ with state $\hat{x}(k),$ control inputs $u(k)$ as:
\small
\begin{equation}
\label{eqn:MPC}
\begin{aligned}
    &\{\Tilde{u}_*^{k+l}\}_{l=1}^N = arg \min_{ \{\Tilde{u}^{k+l}\}_{l=1}^N }~~~ \sum_{l =k+1}^{k+N} \big[ P_c^l+P_h^l +L_k^l \big]\\
    & \ \ \ \ \ s.t.~~ (\{\Tilde{u}^{k+l}\}_{l=1}^N, \{\Tilde{x}^{k+l}\}_{l=1}^N ) \in \Omega, 
\end{aligned}
\end{equation}
\normalsize
where the feasible set $\Omega$ imposes the system dynamics and constraints defined in (\ref{eqn:cont_time}), and $L_k^l = \sum_l \| \Tilde{u}^{k+l}-\Tilde{u}^{k+l-1}\|_R^2$ is a regularizer term, which penalizes large changes in the control inputs to avoid undesirable oscillations. Here, $\Tilde{u}^k = u(k)$ and the
regularization parameter $R$ is tuned to obtain the desired smoothness in the control input trajectory $\{ \Tilde{u}^l\}.$

For the duration of control, the MPC algorithm is computed as follows:
\begin{enumerate}
    \item step 1: Given the state $x^k$ at time $k,$ MPC solves for the optimal $\{u^{k+l} \}_{l=1}^N$ such that (\ref{eqn:power}) is minimized subject to the aforementioned constraints. 
    \item step 2: $u^{k+1}$ is utilized to obtain $x^{k+1}$ from (\ref{eqn:gov_ZoneTemp}a)
    \item step 3: $k=k+1$ and repeat steps $1$ and $2$.
\end{enumerate}

\subsection{Adaptive MPC}
Putting the two previous sections together, we now propose an algorithm for recursive identification of the proposed model for model predictive building control. The algorithm consists of iterative implementation of MHE and MPC. Algorithm 1 outlines the detailed steps  of our method. 
\begin{algorithm}\label{alg:1}
\caption{Algorithm for online model identification and MPC }
\textbf{Input:} time $k,$ $\boldsymbol{x}_o^k=[\hat{x}^{(k-M-1)^T}, \cdots, {\hat{x}^{(k-1)^T}}]^T,$ $\hat{p}^{k-1}, $ $\{y(l)\}_{l=k-M}^{k}, $ $\{u(l) \}_{l=k-M}^{k}$ \\
\textbf{Output:} $u(k+1)$
\begin{algorithmic}[1]
\ForAll{time $k$} \label{step1_start}
\State Using $\boldsymbol{x}_o^k, \hat{p}^{k-1},$ $\{u(l) \}_{l=k-M}^{k},$ $\{y(l)\}_{l=k-M}^{k},$ solve (\ref{eqn:MHE}) to obtain $\boldsymbol{x}_o^{k+1}$ and $\hat{p}^k$
\State Set $\hat{x}(k)= \boldsymbol{x}_o^{k+1}[M], p(k)= \hat{p}^k$
\State $\Tilde{u}^{k} = u(k)$
\State Solve (\ref{eqn:MPC}) to obtain $\{\Tilde{u}_*^{k+l} \}_{l=1}^N$
\State $u(k+1) = \Tilde{u}^{k+1}_*$
\EndFor \label{step1_end}

\end{algorithmic}
\end{algorithm}

\section{RESULTS AND DISCUSSION}

We verify the performance and effectiveness of the proposed estimation and control methodologies in simulation. The ground truth zone models are adopted from \cite{blum2019prototyping} and are simulated using IDA solver. We use June Florida weather at once per five minutes frequency, alongside occupancy profiles that are shown in Figure~\ref{fig:Flowrates}. It is important to note that our proposed algorithm is not limited to this specific type of system but can rather be scaled and adopted for other type of HVAC systems. The main results are threefold: (1) demonstration of the estimation results for MHE; (2) demonstration of the predictive accuracy of identified model; and (3) a comparative analysis between our proposed HVAC control, which we refer to as learning-based MPC (LBMPC), with exact MPC (EXMPC) using noise free dynamics of the building zone and HVAC system, and the following rule-based controller (RBC): similar to deadband control, RBC obeys the following rules: (a) The cooling unit PI tracks a desired supply air temperature setpoint. (b) The PI control of the heating units is activated every time the heating setpoint (lower band) is  violated. (c) The zone PI controller remains engaged until the zone temperature reaches the cooling (upper band) setpoint at which point it disengages until the next violation. During unoccupied hours, the deadband between cooling and heating setpoints is widened and the cooling supply temperature setpoint is increased. These relaxed constraints are often referred to as “setback” which reduces the energy consumption during these unoccupied hours. These RBC rules are similar to control rules deployed in commercial buildings. Both EXMPC and LBMPC are supervisory controllers that manipulate lower level controllers via cooling and heating setpoints $T_{cs}, T_{zs,i}$ with the objective of consuming least amount of energy to keep the occupants satisfied. Before discussing the results in detail, it is worth mentioning that to solve the dynamical system in (\ref{eqn:cont_time}a-\ref{eqn:cont_time}f) for the moving horizon estimation in (\ref{eqn:MHE}) and the MPC in (\ref{eqn:MPC}), a direct collocation with $4^{th}$ order polynomials with Radue collocation points in CasADi \citep{andersson2019casadi} are exploited. Both the estimation and control problems are solved with gradients and hessian approximates using automatic differentiation (AD) in CasADi, with a call to IPOPT \citep{wachter2006implementation} solver. This speeds up the optimization and helps achieve real time performance.
\begin{figure}
   \begin{subfigure}{}
   \centering
   \includegraphics[width=0.9\columnwidth]{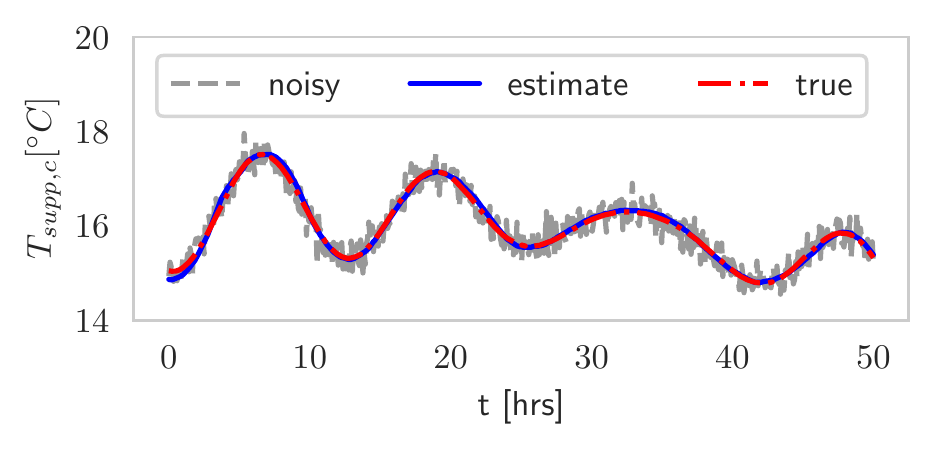}
\end{subfigure}

\begin{subfigure}{}
   \centering
   \includegraphics[width=0.9\columnwidth]{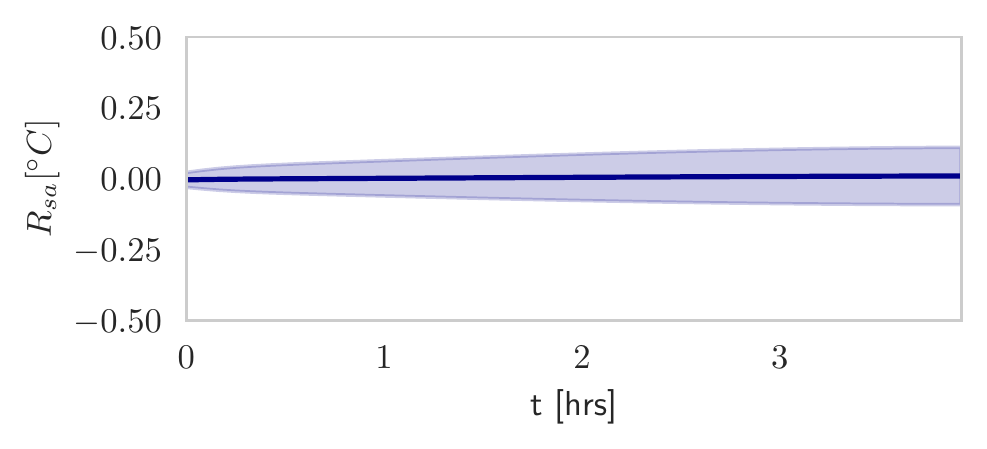}
\end{subfigure}
  \caption{The set of figures show the results of state estimation from noisy measurements of cooling supply air temperature $T_{supp,c}$ (top) and the predictive accuracy of the proposed neural network in (\ref{eqn:fnn}) for a prediction horizon of 4 hours (bottom). For estimation, multisine reference signals are passed to PI controllers in the closed loop simulation with measurements noise $v_k \sim N_{iid}(0,0.2)$ added to all temperatures to generate the ground truth data. As for the predictions, the uncertainty in the plots comes from more than 3000 separate sets of predictions. The solid lined are means and the highlighted areas represent one standard deviation for the residuals. Note that model is not trained on this validation data.}
   \label{fig:estimation}
\end{figure}
\subsection{State estimation and learning evaluation:}
In total, we have 6 neural networks, one for cooling and five for heating units in (\ref{eqn:fnn}) and (\ref{eqn:fnn2}). In all cases, we assume a single hidden layer with three hidden neurons. Since each network has three inputs and a single output, the total number of weights and biases amounts to 96 parameters. In addition, we also have 57 parameters in zone model from (\ref{eqn:gov_ZoneTemp}), for a total number of 153 parameters to be estimated. Figure~\ref{fig:estimation} demonstrate the results of state estimation and model prediction with details provided in the caption. Two things should be noted here: (1) the first plot show estimation post convergence. Empirical evidence suggests that with proper tuning of matrices in (\ref{eqn:MHE}), convergence can be achieved in less than 1000 iterations. (2) since the proposed estimation algorithm is recursive, model parameters are updated, and varying, at every iteration. This can raise questions regarding potential overfitting. To avoid overfitting, we tune $P_p$ to penalize parameter oscillation for a smooth adaptation. Furthermore, it should be noted that, unlike traditional unconstrained gradient descent based training of neural networks, the learning process is constrained in accordance with equation~(\ref{eqn:cont_time}). The bottom figure demonstrates the predictive ability of our modeling approach. Note that since the algorithm itself is recursive, the model is not trained on the future data points and these results should be treated similarly to out of training validation.

\subsection{Control results and discussion:} \label{subsec:discussion}

\begin{figure}[t]
\begin{subfigure}{}
\centering
\hspace{2.5mm}
\includegraphics[width=.9\columnwidth]{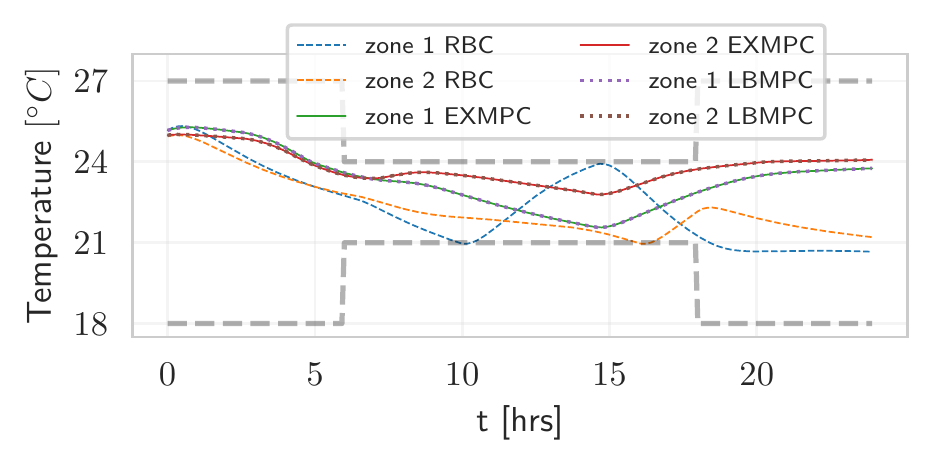}
\end{subfigure}

\vspace{-12.1mm}

\begin{subfigure}{}
\centering
 \includegraphics[width=0.9\columnwidth]{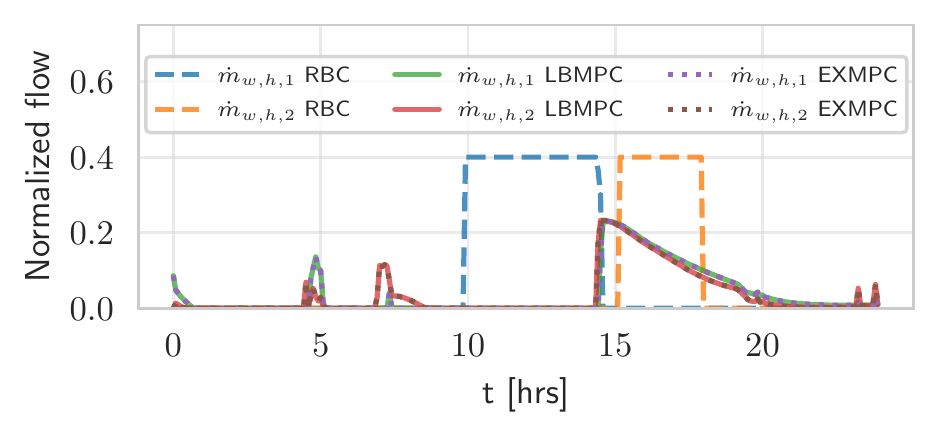}\\
\end{subfigure}
\vspace{-11.1mm} 
\begin{subfigure}{}
\centering
\includegraphics[width=0.9\columnwidth]{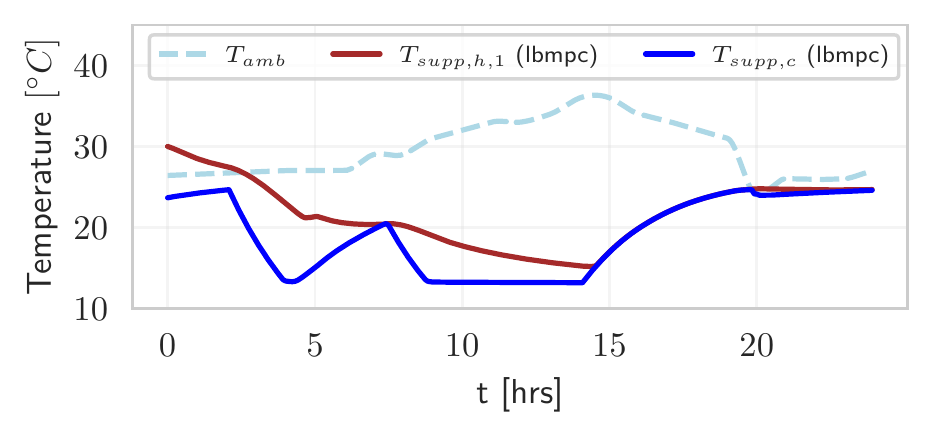}\\
\end{subfigure}
\vspace{-11.1mm} 
\begin{subfigure}{}
\centering
\includegraphics[width=0.9\columnwidth]{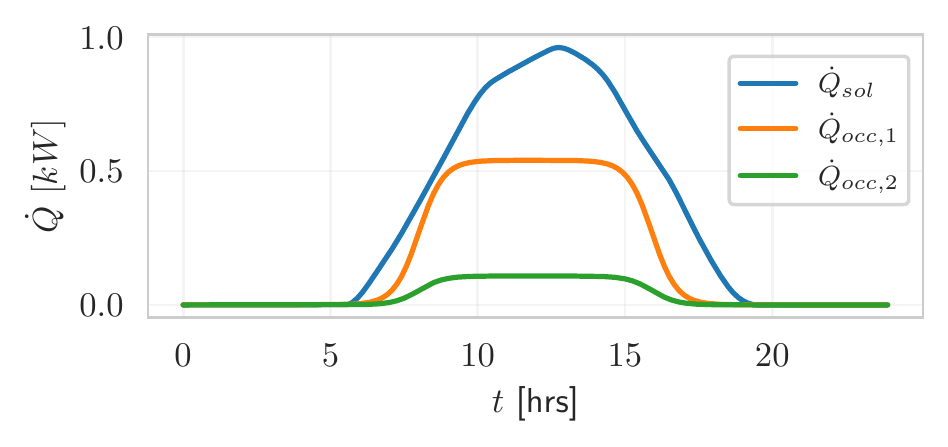}\\ 
\end{subfigure}
\caption{Comparison between RBC, LBMPC and EXMPC Zone temperatures (top), Heating coil flow rate, scaled down by maximum flow rates (second), ambient air temperature, temperatures of the air at HVAC outlet, and the VAV outlets (third), and solar and occupant loads (bottom).  
\label{fig:Flowrates}}
\end{figure}

\textit{Energy savings:} The prediction horizon for both MPC algorithms is 48 data points (4 hours). As expected, and compared to RBC, LBMPC cuts the cooling and heating costs by a noticeable amount during this period of performance. The numbers are as follows: $9.33 \%$ reduction in heating coil energy, $23.2 \%$ in cooling coil energy, and overall there is an energy savings of $18.08 \%$ for the whole system. Most of the energy savings come from cutting overcooling and reheating compared to RBC cooling and heating units respectively. Looking at Figure~\ref{fig:Flowrates}, it is evident that LBMPC is matching the performance of EXMPC. This result is particularly encouraging since it has been achieved with no knowledge of the model and starting from random initialization of parameters until convergence is achieved. It is also somewhat expected from the identification results in Figure~\ref{fig:estimation}.

\textbf{Limitations, challenges and future Work:} A few points should be noted here: in identification, since we do random initialization of parameters, convergence will be achieved over several iterations and most likely during control operation in a real scenario (which can be assumed to deploy a rule-based control policy). In simulation (Monte Carlo convergence) we have observed that with the right choice of $P_p$, convergence happens in less than 1000  iterations (almost 3.5 days of training given 5 minute sampling time) for sufficiently exciting inputs such as chirp and multisine signals. To achieve reasonable performance in practice, there are two possible ways to curtail this problem: (1) use an offline training period (with historical data) and use the results as initialization for faster convergence and (2) Learning during weekend periods where the schedule allows more leeway for excitation. Ultimately, a mathematical proof of convergence is necessary for performance guarantee. We are currently working on this problem. In control, we minimize power consumption of cooling (heating) units and model it as a function of cooling (heating) load of the air. While this is a reasonable model, it has its limitations. An example of this is present in Figure~\ref{fig:Flowrates} where at 8 PM, and during unoccupied hours, a sudden cooling happens to attain thermal equilibrium in equation~(\ref{eqn:power}a). While this is valid for the optimization problem in equation~(\ref{eqn:MPC}), it is counter intuitive to the overall energy savings goal. To curtail this, we are working on a predictive model for flow rates to directly include an energy cost based on electric power consumption in the MPC objective.
\section{CONCLUSION}
In this work, we have successfully demonstrated the modeling and use of nonlinear MHE for estimation and identification and MPC for building with unknown HVAC dynamics. We simultaneously learn the dynamics of the supply air temperatures for HVAC using a neural network and a lumped RC representation of zone level heat transfer. The online estimation and control algorithms have shown encouraging results for identifying a suitable model for MPC in one case study. We have also shown that the proposed algorithm is capable of performing energy saving tasks such as preheating and precooling.
By modeling in CasADi, we leverage algorithmic differentiation to speed up the nonlinear optimization by providing gradients to open source solvers., e.g. IPOPT. To improve our results, we are actively working on further speeding up the optimization by quadratic programming (QP) reformulation of MPC problem and we are also considering forecast uncertainties (non-gaussian in the case of occupants) in a stochastic version of our proposed algorithm. 


\bibliography{ifacconf}             
                                                   







\end{document}